ADVANCED
FUNCTIONAL
MATERIALS



**A memristive nanoparticle/organic hybrid synapstor for neuro-inspired computing.**

By *F. Alibart, S. Pleutin, O. Bichler, C. Gamrat, T. Serrano-Gotarredona, B. Linares-Barranco and D. Vuillaume\*.*

[\*]     Dr. F. Alibart, Dr. S. Pleutin, Dr. D. Vuillaume
Institute for Electronics Microelectronics and Nanotechnology (IEMN), CNRS, University of Lille, BP60069, avenue Poincaré,
F-59652cedex, Villeneuve d'Ascq (France).
E-mail: dominique.vuillaume@iemn.univ-lille1.fr
        O. Bichler, Dr. C. Gamrat
CEA, LIST/LCE (Advanced Computer technologies and Architectures), Bat. 528, F-91191, Gif-sur-Yvette (France).
        Prof. T. Serrano-Gotarredona, Prof. B. Linares-Barranco
Instituto de Microelectrónica de Sevilla (IMSE), CNM-CSIC, Av. Americo Vespucio s/n, 41092 Sevilla (Spain).

[\*\*]    This work was funded by the European Union through the FP7 Project NABAB (Contract FP7-216777). We thank D. Guérin, K. Lmimouni, S. Lenfant (CNRS-IEMN) for help and advises during the device fabrication, and D. Querlioz (CNRS-IEF) for helpful discussions. Supporting Information is available online from Wiley InterScience or from the corresponding author.



**Abstract**.

A large effort is devoted to the research of new computing paradigms associated to innovative nanotechnologies that should complement and/or propose alternative solutions to the classical Von Neumann/CMOS association. Among various propositions, Spiking Neural Network (SNN) seems a valid candidate. (i) In terms of functions, SNN using relative spike timing for information coding are deemed to be the most effective at taking inspiration from the brain to allow fast and efficient processing of information for complex tasks in recognition or classification. (ii) In terms of technology, SNN may be able to benefit the most from nanodevices, because SNN architectures are intrinsically tolerant to defective devices and





performance variability. Here we demonstrate Spike-Timing-Dependent Plasticity (STDP), a basic and primordial learning function in the brain, with a new class of synapstor (synapse-transistor), called Nanoparticle Organic Memory Field Effect Transistor (NOMFET). We show that this learning function is obtained with a simple hybrid material made of the self-assembly of gold nanoparticles and organic semiconductor thin films. Beyond mimicking biological synapses, we also demonstrate how the shape of the applied spikes can tailor the STDP learning function. Moreover, the experiments and modeling show that this synapstor is a memristive device. Finally, these synapstors are successfully coupled with a CMOS platform emulating the pre- and post-synaptic neurons, and a behavioral macro-model is developed on usual device simulator.

## 1. Introduction

Spike-Timing Dependent Plasticity (STDP) is widely believed today to be one of the fundamental mechanisms of the unsupervised learning in biological neural networks. STDP in biological systems is a refinement of Hebb's learning rule.[1] Grant et al.[2], Markram et al.[3], Bi and Poo [4] observed STDP in biological synapses. The principle of STDP is to tune the response of a synapse as a function of the pre- and post- synaptic neurons spiking activity - Fig. 1-a. Depending on the correlation or anti-correlation of the spiking events of the pre- and post-synaptic neurons, the synapse's weight is reinforced or depressed, respectively. The so-called "STDP function" or "STDP learning window" is defined as the relationship between the change in the synaptic weight or synaptic response versus the relative timing between the pre- and post-synaptic spikes (Fig. 1-b).[5] The implementation of STDP with nanodevices is strongly driven by a bio-inspired approach to enable local and unsupervised learning capability in large artificial SNN in an efficient and robust way. To this end, it is envisioned to use the nanodevices as synapses and to realize the neuron functionality with CMOS. This





approach is supported by the fact that the limiting integration factor is really the synapse density, as realistic applications could require as much as $10^3$ to $10^4$ synapses per neuron. Snider [6] proposed an implementation of STDP with nanodevices, where the synapses are realized with a crossbar of memristors [7] and the neurons with a "time-multiplexing CMOS" circuit. Using these two elements, it should be possible to reproduce exactly the "STDP learning window" of a biological synapse (Fig. 1-b). Linares-Barranco et al. simulated the implementation of the STDP function with memristive nanodevices.[8,9] Using a specific shape of the spikes and the non-linearity of the memristor, they showed that the conductivity of the memristor can be tuned depending on the precise timing between the post-synaptic and pre-synaptic spikes. More interestingly, they showed that the shape of the STDP learning window can be tuned by changing the shape of the spike (Fig1-c). We have to emphasize that our aim is to be inspired by the behavior of a biological synapse for neural computation applications (and not to build a model system of the synapse), thus the important point is to reproduce qualitatively the STDP behavior, even if the spike signals applied to the synapstor are not close to the real biological spike.

We recently demonstrated that the Nanoparticle-Organic Memory FET (NOMFET) is able to mimic the short-term plasticity (STP) behavior of a spiking biological synapse.[10] When a sequence of voltage pulses is applied across the device, the current transmitted by the NOMFET is modulated depending on the frequency of the pulses and the past input activity of the device,[10,11] mimicking the facilitating or depressing behavior of a biological spiking synapse.[12] Research on artificial synapse devices mimicking the plasticity of a biological synapse is a burgeoning field. Recently, Jo et al.[13] have observed STDP in Ag/Si-based memristor, Lai et al.[14] in polymer/Si nanowire transistor, Seo et al.[15] in oxide resistive memory, Kuzum et al. in phase-change memory.[16] Here, we demonstrate the STDP behavior





of the NOMFET. First, we carefully analyze the behavior of this synapstor and show that it can be modelized by the memristor equations.[17,18] Thus, we follow the Linares-Barraco et al. suggestions [8,9] to successfully implement the STDP behavior with the NOMFET. Beyond the demonstration at a single device level, we also demonstrate that the NOMFET can be efficiently coupled with a CMOS platform emulating the pre- and post-synaptic neurons. Finally, we developed a behavioral macro-model suitable for device/circuit simulations using commercially available simulators (Spectre-Cadence).

## 2. The NOMFET: a memristive device.

The NOMFET is based on a standard bottom gate/bottom source-drain organic transistor with gold nanoparticles (NPs) fixed at the gate dielectric/organic semiconductor (OSC) interface by surface chemistry (see Experimental section, and a detailed material characterization in Ref. [10]). The STP behavior of the NOMFET is due to the internal charge/discharge dynamics of the NP/OSC system with typical time constants that can be adjusted between 1 to $10^2$ s.[10] While we have demonstrated some simple neuro-inspired plasticity for NOMFETs with a channel length L down to 200 nm, and NP diameter of 5 nm, working at a nominal bias of –3V,[10] here for the sake of demonstration, all the experiments are reported for L = 5 µm NOMFETs and 20 nm diameter NPs working at a nominal voltage of -30V, because these devices previously showed the largest plasticity amplitude (i.e. the largest modulation of the NOMFET output current, here analogous to the synaptic weight, by the applied spike sequence).[10] The channel width (W) is 1,000 µm for the 5 µm length NOMFE, to maximize the output current, given the relative low mobility of the device (ca. $10^{-3}$ $cm^2V^{-1}s^{-1}$).[10] Optimization of the OSC properties (not done here) will allow reaching a state-to-the art mobility of about 1 $cm^2V^{-1}s^{-1}$, and will allow reducing the actual width by a factor $10^3$.





Further optimization would be the use of high-k dielectric to reach the same output current while downscaling W accordingly. Downscaling the NOMFET channel length to 30 nm (with 5 nm diameter NPs) is possible (we have already demonstrated a 30 nm channel length OFET[19]), but such a task would require a hard work for technological optimization, out of the scope of this proof of principle demonstration.

The NOMFET is used as a pseudo two-terminal device (Fig. 2-a): the drain (D) and gate (G) electrodes are connected together and used as the input terminal of the device, and the source (S) is used as the output terminal (virtually grounded through the ampmeter). To establish that it works as a memristive device, we write the output current - input voltage relation in the NOMFET according to the formalism proposed by Chua,[17] and we discuss the significance of the terms in this equation:

$$I_{DS}(t) = G(Q_{NP}(t), V_{DS}(t), t) V_{DS}(t) \qquad (1)$$

$$\dot{Q}_{NP}(t) = g(Q_{NP}(t), V_{DS}(t), t) \qquad (2)$$

where $G$ is the conductance of the device that includes the field effect, $V_{DS}(t)$ is the applied signal of time varying shape, and $Q_{NP}(t)$ the charges trapped in the NPs. For the NOMFET, $Q_{NP}(t)$ is the relevant internal parameter, and its first-order time derivative is given by the $g$ function, which is the "memristive" function that describes how this internal parameter is updated as function of the internal state, the external voltage and time. A non-linear behavior of $g$ is very interesting to implement synaptic plasticity and STDP.[6,8,9,18] A $g$ function with a null value between negative and positive threshold voltages and increasing/decreasing parts above/below (respectively) these thresholds has been used to simulate STDP and learning capabilities in memristor-based neuro-inspired circuits.[8,9]

To characterize the memristive behavior of the NOMFET, we measure the change of its internal parameter $\delta Q_{NP}$ when voltage signal $V_{DS}(t)$ is a pulse of amplitude $V_P$ and duration





10 s. This value of 10 s has been fixed in order to maximize the effect of the NP charge. This time is longer than the typical charging/discharging time constants (about 2-3 s)[10] for a NOMFET with a channel length of 5 μm and 20 nm NPs used for these experiments. Reducing the width of the charging pulse will give smaller variations of the current, but does not change the conclusions. The output current, before ($I_{Initial}$) and after ($I_{after}$) the application of the charging pulse, are measured with a short read pulse (100 ms). This pulse is short enough to not modify the charge state of the NPs. Plotting *($I_{after} - I_{initial}$)/$I_{initial}$*, which is proportional to $\delta Q_{NP} = Q_{NP}^{After} - Q_{NP}^{Initial}$ (Eq. S24, supporting information), versus $V_P$ gives a representation of the g-function of the NOMFET. As the current at a given time *t* depends on the history of the device, we have developed a specific reset protocol (see Experimental section, and Fig. S1, supporting information) that sets the charge state of the NPs to the same $Q_{NP}^{Initial}$ before each measurement at different $V_P$. Figure 2-c shows the measured relative variation of the current (red dots) as a function of $V_P$, i.e. the internal memristive-like function of the NOMFET. This function displays the three expected regions similarly to the resistance change in a voltage-controlled memristance:[8,9,17,18] (i) For the negative voltage, the NPs are charged with holes, the Coulomb repulsion between the positively charged NPs and the OSC reduces the hole density in the conducting channel, the conductivity of the NOMFET is decreased. (ii) For intermediate voltages ($V_{th1} < V < V_{th2}$), the effect of the input voltage on the charge state of the NPs is null. The charge state of the NPs cannot be changed. The physical meanings of the two threshold voltages, $V_{th1} \approx 0$ V and $V_{th2} \approx 15$ V, are discussed in the supporting information. (iii) For large positive voltages, holes can be detrapped from the NPs, leading to a reverse effect, i.e. an increase in the conductivity of the NOMFET. The memristive *g* function shown in Fig. 2-c can be calculated using Eqs. S31 (see supporting information) considering the three parts of the experimental curve. For simplicity, we assume





the same time constants in Eqs. S31 ($\tau = \tau_0 = \tau_+ = \tau_- = 5$ s). This value is in good agreement with experimental values for the NOMFET.[10,11] The blue squares in Fig. 2-c are the fit of this model. Eq. S31 gives two linear relationships for the two branches that fit relatively well our data.

### 3. STDP behavior of the NOMFET

In Ref. 10, the STP (short-term plasticity) is obtained by virtue of the unbalanced charging (during the application of a pulse at the input terminal) and discharging (between two successive pulses at the input terminal) behaviors of the NPs, respectively. Here, as detailed below, we play with the same charging/discharging dynamics to modulate (i.e. increase or decrease) the amount of charges trapped in the NPs when two pulses are now applied, one at the input and one at the output terminals of the NOMFET separated by a given time interval, leading to the long-term depression (LTD) or long-term potentiation (LTP) behavior of STDP, respectively. More precisely, figures 3-a shows the two different shapes of the spikes that are applied to the NOMFET, in agreement with the spike shape suggested in Ref. [8] (Fig. 1-c). These spikes are designed so that – when applied alone – they do not induce any significant variation of conductivity. It means that NPs charging and discharging are well balanced between the negative and positive parts of the spike, respectively. The integral of the negative part of the signal ($V < V_{th1} \approx 0$V) is equal to the integral of the positive part $V > V_{th2} \approx 15$V) – Fig. 3-a. To facilitate the measurement with the probe-station, the post-synaptic spike (that must be applied to the $V_S$ terminal of the device) is inverted and applied to the $V_D$ pre-synaptic terminal. Thus, the effective signal (Fig. 3-b) applied to the $V_D$ terminal becomes equivalent to the application of the pre-synapse spike at the $V_D$ terminal and the post-synapse spike at the $V_S$ terminal (as a feedback). We check in section 4 that this procedure gives the





same results as if we had applied the pre- and post-synaptic pulses directly to each of the two terminals. Note that the post-synaptic shape (Fig. 3-b) is slightly different from the pre-synaptic one to take into account the asymmetry of the memristive $g$ function of the NOMFET (Fig. 2-c).

In a first stage, the pre-synapse spike is applied alone at the input terminal of the NOMFET. This step is crucial to verify that the pre-synapse spike alone does not change the conductivity of the NOMFET. In a second stage, we apply the pre- and post-synaptic spikes with a fixed time shift $\Delta t$ between them (Fig. 3-b). The spikes have a frequency of 0.1 Hz and the conductivity of the NOMFET is read with a short pulse (100ms) synchronized with the spike sequence and applied 1s after the end of the pre-/post-synapse spike sequence (Fig. 3-b). The superposition of the pre- and post-synaptic spikes leads to an effective voltage across the NOMFET (bottom Fig. 3-b) in which the positive and negative contributions are no longer equal. This unbalanced contribution allows reproducing the basic principle of the STDP (Fig. 3-c). (i) When the pre-synaptic neuron fires alone, the weight of the synapse is not changed. In the first part of figure 3-c (labeled "No post-spike"), 10 pre-synaptic spikes are applied alone to the NOMFET. The conductivity of the NOMFET remains in its initial state. (ii) When a pre-synaptic spike is correlated with a post-synaptic spike, the conductivity of the NOMFET is increased (figure 3-c labeled "With post-spike", $\Delta t = + 2$ s, 13 correlated spikes) due to the more important contribution of the positive part of the effective voltage across the NOMFET (i.e. the NPs are progressively discharged). The synaptic weight is reinforced. (iii) When the post- and pre-synaptic neuron spikes are anti-correlated ($\Delta t = - 2$ s), the conductivity decreases, the contribution of the negative potential part dominates and the NPs are gradually more charged. The weight of the synapstor is depressed.

The same data are plotted as $\Delta I/I$ vs. $\Delta t$ curves (STDP learning curve) in figures 4-a and 4-b for a sequence of 12 successive triangular and square spikes, respectively. Fig. 4-a





(triangular spike) qualitatively looks like the one reported by Bi and Poo [4] for a biological

synapse, by other groups with inorganic devices [13-16] and Linares-Barranco et al.[8,9] for

simulations on memristors, i.e. a more or less sharp STDP function as shown in Fig. 1-c (right

upper corner). Results in Fig. 4-b obtained with a rectangular spike show that the shape of the

STDP learning window can be modulated successfully by changing the shape of the pulses, in

good agreement with the behavior predicted by Linares-Barranco for a memristive device [8,9]

(Fig. 1-c). Recent results on synapses based on phase change memory also showed

experimentally that it is possible to change the shape of the STDP curves, albeit with a much

more complicated sequence of spikes in this case. [16] Now, we obtain a more "squared" or

"rounded" shape for the NOMFET STDP function, comparable with the simulation (right-

lower corner in Fig. 1-c). Our model reproduces the experiments with a good qualitative

agreement (blue squares figure 4) considering five different values for the charge/discharge

time constants depending on the voltage (Eqs. S32 and S34, supporting information). These

time constants $\tau_i$ (-2 < i < +2, Eq. S34) are in the range 0.3 to 5 s, in good agreement with

previous measurements showing that the charging/discharging of the NPs follows a multi-

time constant dynamics in this time-scale range (Fig. S5 in the supporting information of Ref.

[10]). Finally, we can note that the approximation used in Eq. S24 ($\gamma\delta Q_{NP}$ << 1) is justified (see

Fig. 2-c) at low bias and is reasonable for bias voltages in the range ± 30V used in the STDP

experiments. Nevertheless, the model-experiment agreement seems not strongly affected

when $\gamma\delta Q_{NP}$ approaches 1 at higher voltages. Finally, we note that the STDP amplitude (from

– 15% to 30%, Fig. 4) is lower than for biological synapse (-40 to 100%) as reported by Bi

and Poo, [4] however, our results are larger or similar to the ones reported by other groups. [13-

16] We expect that these performances can be improved by a careful technology optimization,





for instance, recent STDP results with phase change memory (PCM) [16] – a much more

mature technology – reached a dynamic between – 40 and 110%.

## 4. Hybrid NOMFET/CMOS system.

Instead of using a single device connected to a probe-station, a more realistic demonstration

of the STDP behavior of the NOMFET is obtained by interfacing these synapstors with a

CMOS-based electronic board to emulate the neurons and generate pre- and post-synaptic

spikes, which are now directly applied to the input and output of several NOMFETs. Several

NOMFETs were mounted in a TO case and plugged on the electronic board (see Fig. S2,

supporting information). This board is driven by an FPGA and is remotely controlled by a PC

(see details in the supporting information). Series of rectangular spikes, identical to those used

for the previous measurements, are applied simultaneously to two NOMFETs, with a

randomly generated time interval $\Delta t$ between the pre- and post-synaptic spikes. The output

currents of these NOMFETs are acquired with the electronic-board (see Fig. S2, supporting

information). The $\Delta I/I$ versus $\Delta t$ measured simultaneously for two NOMFETs are shown in

Fig. 4-c. The STDP function obtained with this NOMFET/CMOS system is in good

agreement with the one measured point-by-point for a NOMFET connected with the probe-

station as shown in Fig. 4-b. In addition this is, to the author's best knowledge, the first actual

implementation of STDP on a dynamic device that meets the following conditions: (i) The

correct behavior is achieved regardless of the initial state of the device, as the timing between

the pre- and post-synaptic spikes is random between each measurement (the same STDP

behavior - Fig. 4-c – has been obtained here with random $\Delta t$, while the data shown in Figs. 4-

a and 4-b have been recorded for a linear sequence of $\Delta t$ from -5 to +5 s). (ii) The behavior

remains consistent and very well reproducible regardless of the characteristics of the devices.

Indeed, there is a factor 10 in the mean conductivity ratio between the two NOMFETs used in





Fig. 4-c and yet the relative change in conductivity is the same for the two devices, i.e. the variability on the dynamical behavior of the NOMEFT is very low. This behavior is due to the fact that the STDP is based on a temporal coding, and only the relative variation of the NOMFET conductivity obtained through the applied pulses, and the natural relaxation of the NPs, impose the dynamics. This means that with STDP, we have a reliable way of programming conductivity changes using temporal information coding with seemingly unreliable devices. As a consequence, STDP and NOMFET can be useful to implement some learning algorithms in neural network circuits without to pay too much attention to some common variability sources, such as physical dimensions, reproducibility and control of the technological steps.

## 5. Behavioral macro-model for neuro-inspired circuit simulation

The physical model developed for such a diode-connected NOMFET (Fig. 5-a) is implemented in SPECTRE-CADENCE for simulating neuro-inspired circuits using STDP and NOMFET. The NOMFET device can be described behaviorally using the macro model circuit shown in Fig. 5-b. The terminal drain and source voltage $V_D$ and $V_S$ are copied to an internal diode in series with a resistor, attenuated by a scaling factor $\alpha$. This is to adapt the operating voltage (few tens of volts) of the NOMFET to a regular silicon diode used in CADENCE. The current through the diode $i_{ds0}$ is sensed and copied to the bottom input of element $m(\ )$. Element $m(\ )$ computes the following function:

$$m(i_{ds0}, w) = A i_{ds0} e^{-w/w_0} \tag{11}$$

where $w$ is a circuit variable (a voltage) that describes the evolution of the charge in the NPs, $w(t) \propto \delta Q_{NP}(t)$ (Eq. S36 in supporting information). Internal voltage $w$ is generated by feeding a resistor $R$ and a capacitor $C$ with a current source of value $-C\rho(V_{DS})$. The time constant in eqs. (S35-S39) is such that $\tau = RC$. This way this circuit implements Eqs. (S38). This macro





model is used to simulate the behavior of the NOMFET when stimulated by a signal such as the one shown in 2-b, a pulse $V_P = $ -35V during 10s. By holding $V_S = 0$ and applying a negative -35V pulse during 10 s at $V_D$, we obtain the signal evolutions shown in Figs 5-c and 5-d. The different parameters were optimized to best fit the measured $I_{DS}$ signal in Fig. 2-b: $\tau$ = 2.2 s ($C$ = 1F, $R$ = 2.2$\Omega$), $A$ = $10^{-6}$, $R_d$ = 20 k$\Omega$, $V_{th}$ = 15V, $w_0$ = 0.16V and $\alpha$ = 0.1. The internal diode is described by $i_{ds0} = I_{d0}e^{V_{diode}/U_T}$ where $U_T = kT/q$ is the thermal voltage ($\approx$ 26mV) and $I_{d0}$ = 8x$10^{-20}$ A. Simulated results in Fig. 5-c are in very good agreement with the experiments (Fig. 2-b). Again, note that the fitted time constant is in good agreement with experimental values for the NOMFET as reported elsewhere.[10] These results validate the macro-model that can be further used to simulate neuro-inspired circuits using STDP learning rules.

## 6. Discussion and Conclusion

Finally, we can notice that the potentiation (depression) reported here for the correlated (anti-correlated) spikes resembles that of a biological synapse (albeit with spike signals adapted to the NOMFET for which the physical mechanisms responsible for the STDP behavior are clearly different from the ones in a biological synapse) as reported by Markram et al.[3] and by Bi and Poo [4], while at different time scales due to the different internal dynamics of the two systems. We have already demonstrated that NOMFET with a smallest channel length (L = 200 nm, and 5 nm NPs), working at a lower voltage (- 3V) exhibit neuro-inspired short-term plasticity (STP) with smaller time constants (~1 s, see Fig. 6-c in Ref. 10), while with a weaker amplitude.[10] So we believe there is room to improve the neuro-inspired behavior of these synapstors and their future use in neuro-inspired computing circuits and architectures. For instance, the actual low time scale response of NOMFET can be ascribed to two features.





(i) The fist one is the low charge/discharge time constants of the NPs, which are capped by alkyl chains (see Experimental section) acting as tunnel barrier. (ii) The relatively low mobility of charges in the pentacene/NP channel (see a discussion in Ref. [10]), which reduces the functioning speed of the device. Improvements (i.e. shorter time-scale, closer to the one of a biological synapse) can probably be attainable by changing the nature of the NP capping molecules (e.g. using more conducting π-conjugated molecules), and/or optimizing the deposition/nature of the organic semiconductor to increase the charge carrier mobility.

## 7. Experimental

*Device fabrication.* The synapstors are made on a highly doped ($\sim 10^{-3}$ $\Omega$.cm) p-type silicon covered with a thermally grown 200 nm thick silicon dioxide. After a usual wafer cleaning (sonication in chloroform for 5 min, piranha solution ($H_2SO_4$ /$H_2O_2$, 2/1 v/v) for 15 minutes - *caution: preparation of the piranha solution is highly exothermic and reacts violently with organics*, ultraviolet ozone cleaning (ozonolysis) for 30 minutes), we evaporated titanium/gold (20/200 nm) electrodes, patterned by optical lithography and lift-off. To attach the NPs, the gold (Au) electrodes were functionalized with a 2-amino ethanethiol molecules (10mg/mL in ethanol) during 5h. After rinse (isopropanol) and subsequent drying in argon stream, the $SiO_2$ surface was functionalized at 60°C during 4min by 3-aminopropyl trimethoxysilane (APTMS) molecules (from ABCR) at 1.25µL/mL (in anhydrous toluene).[20] The reaction took place in a glove-box (MBRAUN) filled with nitrogen (less than 1 ppm of oxygen and water vapor). We removed non-reacted molecules by rinse in toluene, and then in isopropanol under sonication, and the samples were dried under argon stream. This sample was then dipped in an aqueous solution of citrate-stabilized Au-NP (colloidal solution purchased from Sigma Aldrich, 20 ± 3 nm in diameter) overnight under argon atmosphere,





followed by cleaning with deionized water and isopropanol, and drying under argon stream. NP concentration in the solution and duration of the reaction are selected from our previous work to have a NP density on the surface of about $10^{11}$ NP/cm$^2$ that gives the best results for the synaptic behavior of the NOMFET.[10] Then, the Au-NPs were encapsulated by dipping in a solution of 1,8-octanedithiol (from Aldrich) in ethanol (10μL/mL) during 5h. The sample was finally rinsed in alcohol and dried in argon stream. The device is completed by evaporating (substrate kept at room temperature during the evaporation) 35 to 50 nm thick of pentacene at a rate of 0.1 Å/s. More details on the structural characterizations of the NPs networks and pentacene films (SEM, AFM,…) are given in Ref. [10].

*Electrical measurements.* The NOMFET were contacted with a micromanipulator probe station (Suss Microtec PM-5) inside a glove box (MBRAUN) with controlled nitrogen ambient (less than 1 ppm of water vapor and oxygen). Such a dry and clean atmosphere is required to avoid any degradation of the organics. The input spikes were delivered by an arbitrary waveform generator (Tabor Electronics 5062) remote controlled by a PC. The pulse and spike sequences were designed with Matlab. The output currents were measured with an Agilent 4155C semiconductor parameter analyzer.

*Reset protocol.* The reset signal is based on the same principle than the one used to remove the permanent magnetization of a magnet. We impose a decreasing sinusoidal input voltage (see Fig. S1, supporting information) with a large period and a large initial voltage (the period and initial voltage must be large enough in comparison to the input voltage used during the operation/characterization of the device). The NPs are alternatively charged and discharged with a decreasing magnitude. Even if this initial state of charge of the NPs is different from the virgin state of charge of the NPs (i.e. in the as-deposited state), it allows starting a specific measurement from the same initial condition.






[1]     D. Hebb, *The organization of behavior.* (Wiley, 1949).

[2]     K. Gant, C. Bell, V. Han, *J. Physiol. Paris* **1996,** *90*, 233.

[3]     H. Markram, J. Lubke, M. Frotscher, B. Sakmann, *Science* **1997**, *275*, 213.

[4]     G.Q. Bi, M.M. Poo, *J. Neurosci.* **1998**, *18*, 10464.

[5]     J. Sjöström, W. Gerstner, *Scholarpedia* **2010**, *5*, 1362.

[6]     G.S Snider, in *IEEE/ACM International symposium in nanoscale architectures*    85 (2008).

[7]     D.B. Strukov, G.S. Snider, D.R. Stewart, R.S. Williams, *Nature* **2008**, *453*, 80.

[8]     B. Linares-Barranco, T. Serrano-Gotarredona, *IEEE Nano2009*, **2009**, 601.

[9]     C. Zamarreno-Ramos, L. Camuñas-Mesa,J. Pérez-Carrasco,T. Masquelier, T. Serrano-Gotarredona, B. Linares-Barranco,  *Frontiers in Neuroscience* **2011**, **5**, 26.

[10]    F. Alibart, S. Pleutin, D. Guérin, C. Novembre, S. Lenfant, K. Lmimouni, C. Gamrat, D. Vuillaume, *Advanced Functional Materials* **2010**, 20, 330.

[11]    O. Bichler, W. Zhao, F. Alibart, S. Pleutin, D. Vuillaume, C. Gamrat, *IEEE Trans. Electron Devices* **2010**, *57*, 3115.

[12]    J.A. Varela, K. Sen, J. Gibson, J. Fost, L.F. Abbott, S.B. Nelson, *J. Neurosci.* **1997**, *17*, 7926.

[13]    S.H. Jo, T. Chang, I. Ebong, B. B. Bhadviya, P. Mazumder, W. Lu *Nano Letters* **2010**, *10*, 1297.

[14]    Q. Lai, L. Zhang, Z. Li, W. F. Stickle, R. S. Williams, Y. Chen, *Advanced Materials* **2010**, *22*, 2448.







[15]    K. Seo, I. Kim, S. Jung, M. Jo, S. Park, J. Park, J. Shin, K. P. Biju, J. Kong, K. Lee, B. Lee, H. Hwang, *Nanotechnology* **2011**, *22*, 254023.

[16]    D. Kuzum, R. G. D. Jeyasingh, B. Lee, H.-S. P. Wong, *Nano Lett.* **2011**, ASAP on-line, doi:10.1021/nl201040y.

[17]    L.O. Chua, *IEEE Trans. on Circuit Theory* **1971**, *18*, 507.

[18]    M. Di Ventra, Y.V. Pershin, L.O. Chua, *Proceedings of the IEEE* **2009**, 97, 1717.

[19]    J. Collet, O. Tharaud, A. Chapoton, D. Vuillaume, *Appl. Phys. Lett.* **2000**, 76, 1941.

[20]    D.F. Siqueira Petri, G. Wenz, P. Schunk, T. Schimmel, *Langmuir* **1999**, *15*, 4520.






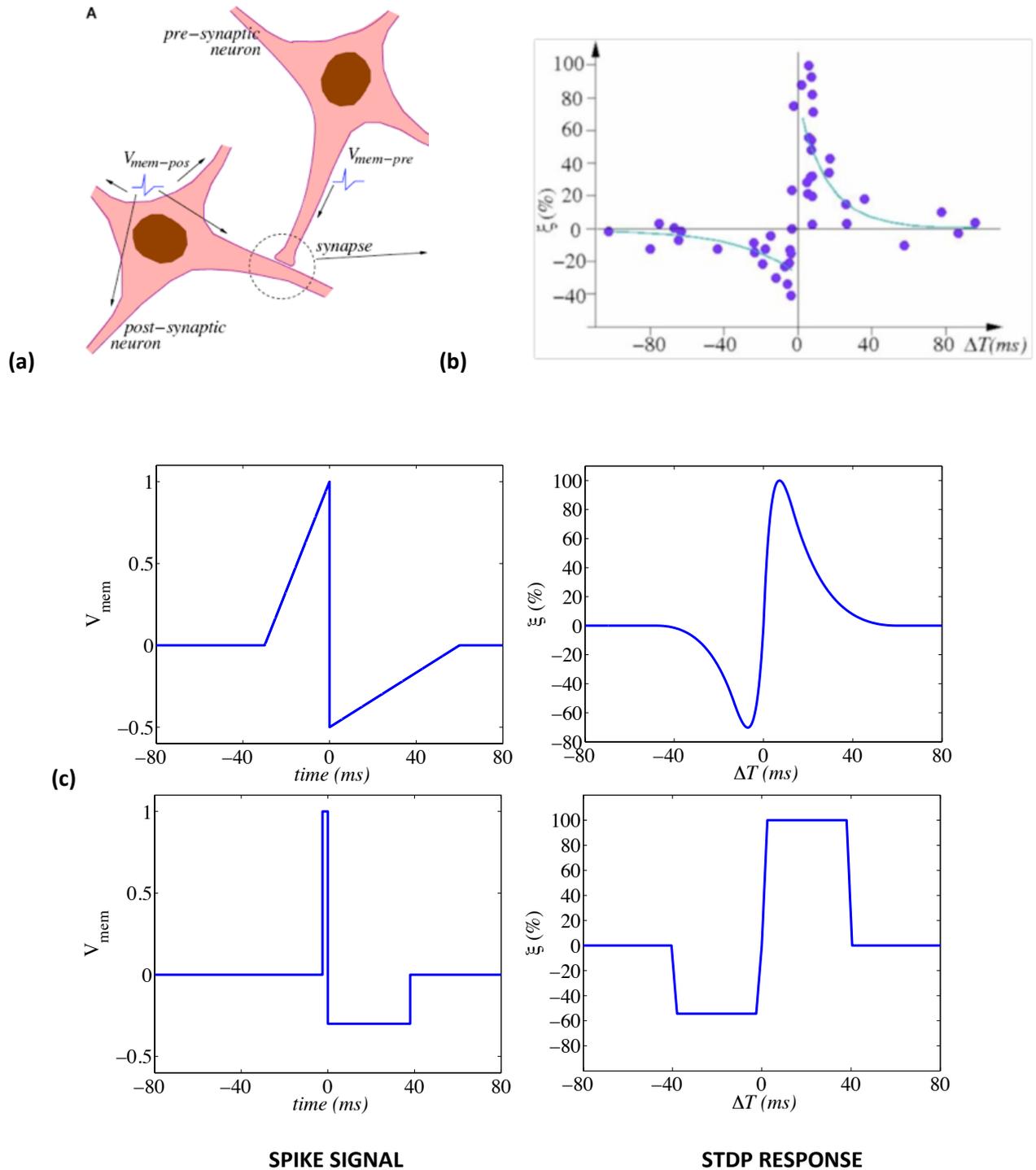

**Figure 1. (a)** Illustration of a synapse connecting two neurons: the pre-synaptic and the post-synaptic neurons (after Ref. [8]). **(b)** STDP function, i.e. change in the synaptic weight versus spike timing interval, measured on a biological synapse (after data from Bi and Poo)[4], **(c)** Two shapes of spikes (left side) and the corresponding STDP functions (right side) calculated for a memristive device (after Linares-Barranco et al.)[8,9]





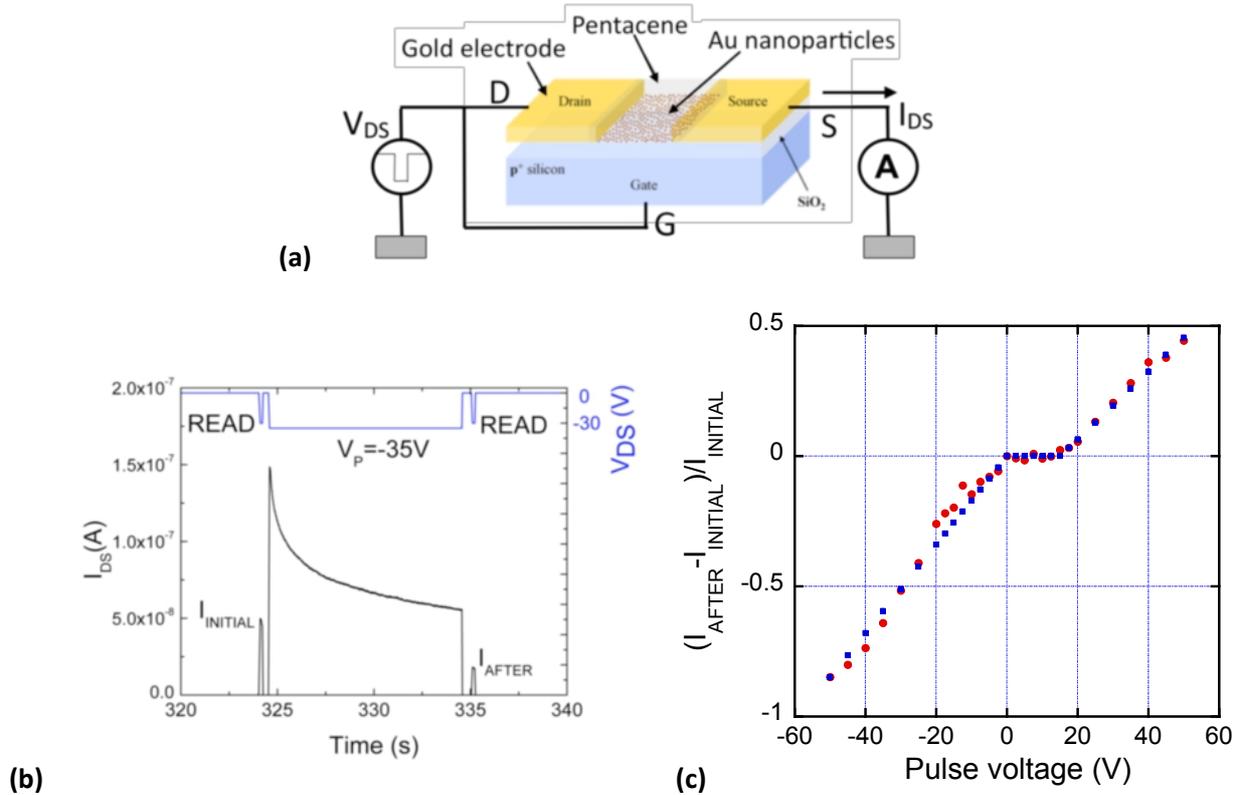

**(a)**

**(b)**

**(c)**

**Figure 2: (a)** Schematic representation of the NOMFET and pseudo two-terminal connections of the device. **(b)** Typical "single pulse measurement" used to characterized the NOMFET as a memristive device. The current is measured just before and after a large pulse of 10s in order to estimate the effect of the large pulse voltage $V_P$ on the NPs charge. **(c)** Relative variation of the current as a function of the pulse voltage $V_P$. Red dots are the experimental measurements and blue squares from the physical model (see supporting information).





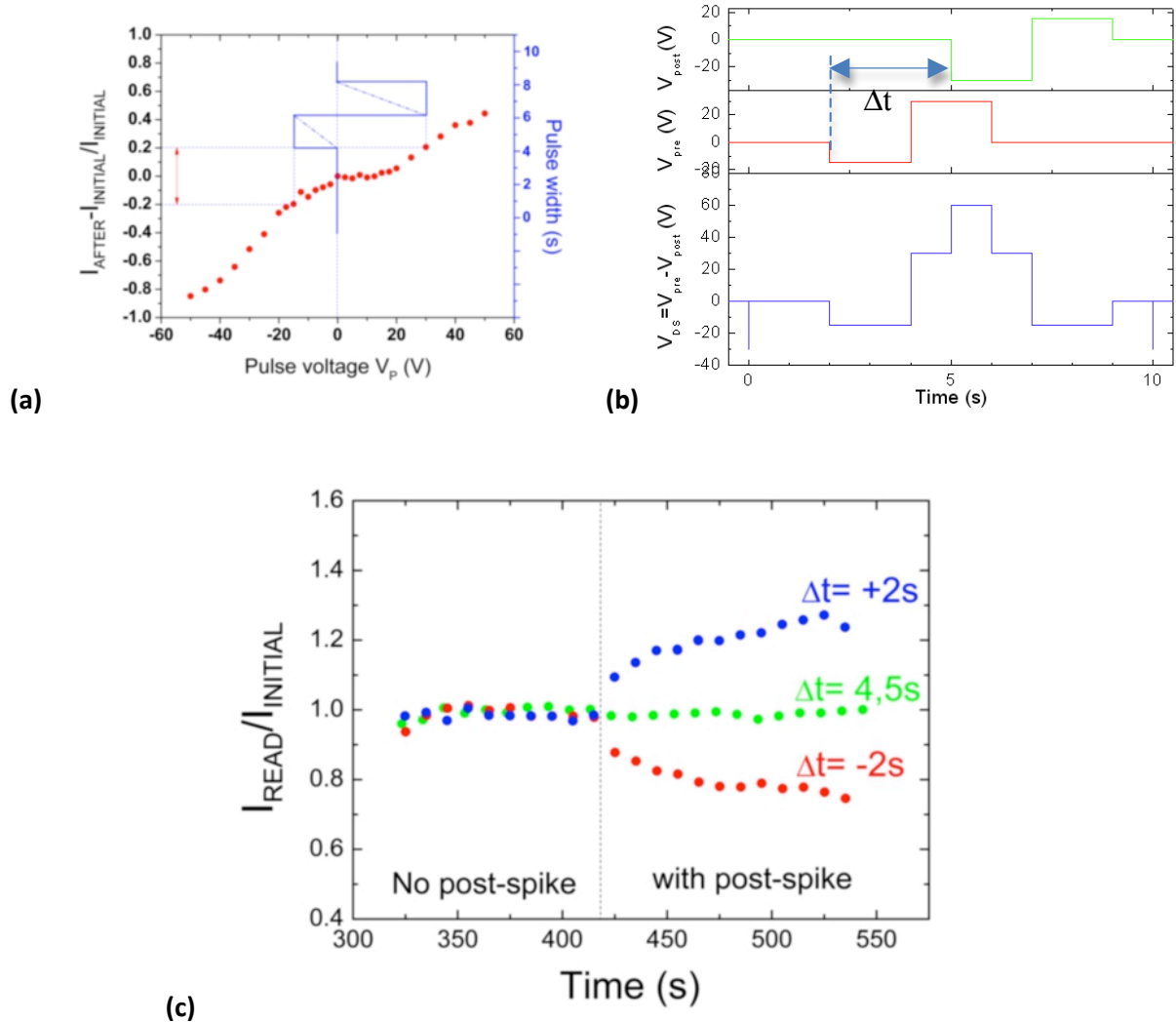

**(a)**

**(b)**

**(c)**

**Figure 3.** **(a)** The two different pulses used to reproduce the STDP: square signal (solid lines), triangular signal (dash-dot line). In the case of the pre-synaptic pulse, the effect of the negative part - $V$ = -15 V for 2s - on the conductivity is equal to the effect of the positive part of the pulse - $V^+$ = 30V for 2 s. **(b)** Pre- (in red) and post-synaptic (in green) pulses superposition: the effective potential across the device is $V_{PRE}$ - $V_{POST}$ (in this case, $\Delta t$ is 3s). Note that the post-synaptic pulse is $V$ = -30 V and $V^+$ = 15V to take into account the asymmetry of the memristive $g$ function of the NOMFET. In this situation, the effect of the post-synaptic pulse alone on the conductivity is null. **(c)** Typical STDP measurement. First, 10 pre-synaptic pulses are applied alone at 0,1 Hz in order to verify that the conductivity is not changed by the pre-synaptic signal alone. Next, 13 pre- and post-synaptic pulses are applied with 3 different $\Delta t$ values.





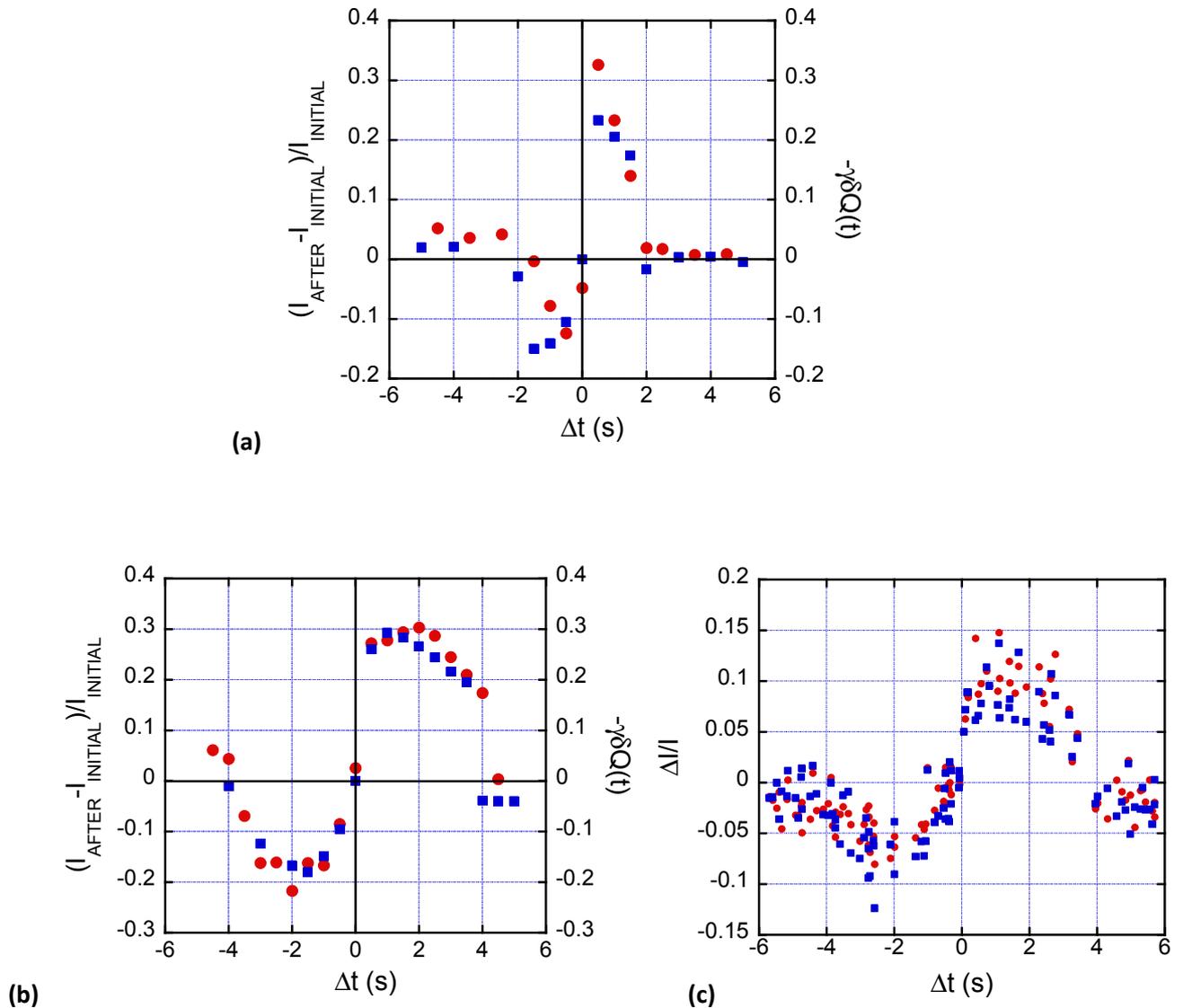

**Figure 4.** The relative variation of current is measured after 12 repetitions of the pre and post pulses pattern with a given *Δt* (as described in figure 3-b). The red dots correspond to the experimental measurement *($I_{After} - I_{Initial}$)/$I_{Initial}$* and the blue squares are the model calculation $-\gamma\delta Q_{NP}$ (see supporting information). **(a)** STDP function obtained with the triangle-shape pulses. **(b)** STDP function obtained with the square-shape pulses. **(c)** STDP learning function acquired with the electronic-board for two NOMFETs (blue and red dots) measured simultaneously.





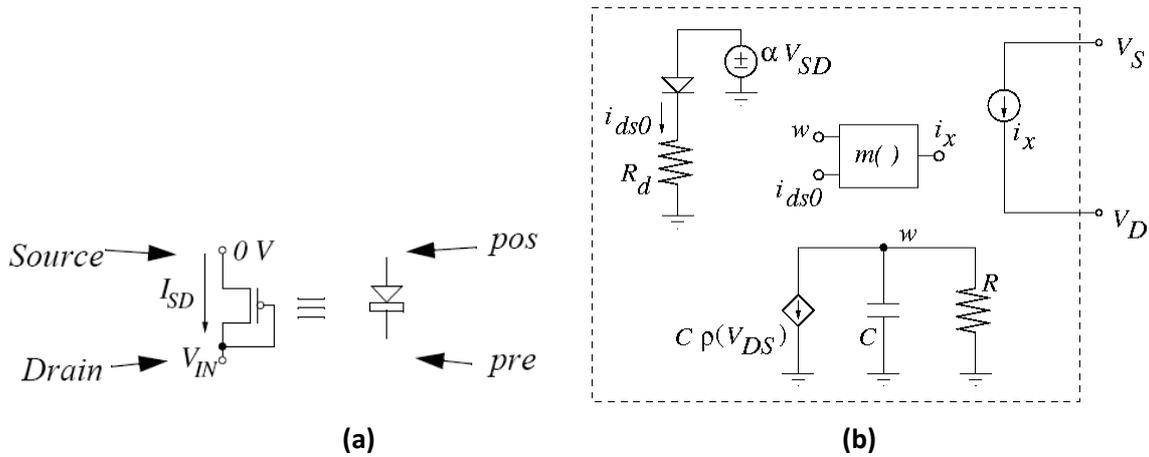

**(a)**

**(b)**

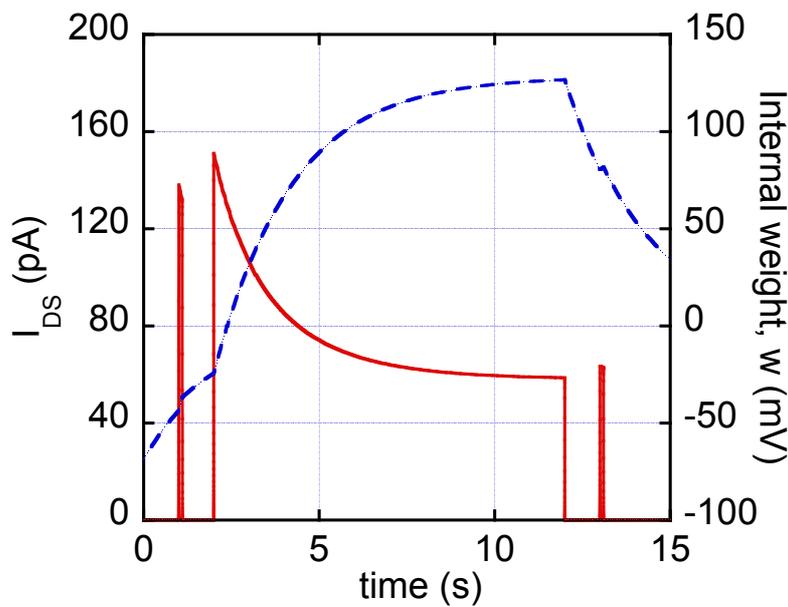

**(c)**

**Figure 5. (a)** The NOMFET is a p-type FET, it is used in a diode-like connected configuration. Source S is the top terminal, drain D is the bottom terminal. $I_{DS}$ is either zero or positive. It is equivalent to a diode. When used as an STDP synapse (see Fig. 2-a), bottom terminal is the pre-synaptic connection and top terminal is the post-synaptic connection. **(b)** NOMFET macro-model implemented in SPECTRE-CADENCE. Simulation of the NOMFET with the macro model: **(c)** output current (solid line, left scale) and evolution of the internal weight parameter $w$ (dashed line, right scale).





**The table of contents entry.** A synapstor (synapse-transistor), called NOMFET (nanoparticle organic memory FET) is designed and fabricated to mimic the spike-timing dependent plasticity (STDP) of a biological synapse. STDP is a fundamental mechanism of learning in the brain. The STDP behavior means that the synaptic response (here the device conductance) depends on the time correlation between pre- and post-synaptic spikes received by the synapstor.




F. Alibart, S. Pleutin, O. Bichler, C. Gamrat, T. Serrano-Gotarredona, B. Linares-Barranco and D. Vuillaume.


A memristive nanoparticle/organic hybrid synapstor for neuro-inspired computing.

ToC figure ((55 mm broad, 50 mm high, or 110 mm broad, 20 mm high))

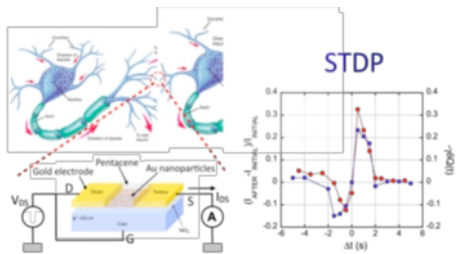

Column Title: *F. Alibart et al.*/Short title





ADVANCED
FUNCTIONAL
MATERIALS

**Supporting Information**.

## A memristive nanoparticle/organic hybrid synapstor
## for neuro-inspired computing.


F. Alibart,[1] S. Pleutin,[1] O. Bichler,[2] C. Gamrat,[2]

T. Serrano-Gotarredona,[3] B. Linares-Barranco[3] & D. Vuillaume.[1]

*(1). Institute for Electronics Microelectronics and Nanotechnology (IEMN),*

*CNRS,*

*Avenue Poincaré, F-59652 Villeneuve d'Ascq, France.*

*(2). CEA, LIST, Embedded Computers Laboratory,*

*91191 Gif-sur-Yvette Cedex, France.*

*(3). Instituto de Microelectrónica de Sevilla (IMSE),*

*CNM-CSIC, Av. Americo Vespucio s/n,*

*41092 Sevilla, Spain.*

*E-mail : dominique.vuillaume@iemn.univ-lille1.fr*


## Supporting information.

**Theory**

1- *Carriers number and source-drain current: effects of trapped charges in the nanoparticles*.
The charge transport in organic thin film is usually interpreted in terms of incoherent hopping of charges via localized states randomly distributed in space and in energy. A particularly simple theory within this line was proposed twenty years ago by Vissenberg and Matters (VM) [2]. The principle merit of their work is to provide simple analytic formula. The important point for us is to describe the changes in the conductivity and in the number of carriers induced by the charges trapped in the gold nanoparticles (NPs). In our previous work [1, 3], we have therefore extended the VM model to take account of these effects. We start to recall the main formula of Refs. [1-3], and then show that the NOMFET enters in the class of memristive devices defined by Chua [4, 5].





Due to disorder an organic thin film can be considered as an electrical network of quantum localized states with hopping transport from site-to-site. The energies of those states, $\varepsilon$ (<0, below the vacuum level), are assumed by VM to be exponentially distributed

$$D(\varepsilon) = \frac{N_t}{k_B \theta_0} \exp\left(\frac{\varepsilon}{k_B \theta_0}\right) \qquad (S1)$$

$N_t$ is the number of localized states per unit volume, $k_B \theta_0$ a measure of the energy width of the distribution ($k_B$, the Boltzmann constant). Each pair of sites, $i$ and $j$, distant by $R_{ij}$ is connected by a bond with conductance $G_{ij}=G_0\exp(-S_{ij})$ where

$$S_{ij} = 2\alpha R_{ij} + \frac{\left|\varepsilon_i - \varepsilon_F\right| + \left|\varepsilon_j - \varepsilon_F\right| + \left|\varepsilon_i - \varepsilon_j\right|}{2 k_B \theta} \qquad (S2)$$

The first term takes account for usual tunnelling processes and the second for thermally assisted tunnelling. $\alpha$ is an effective overlap parameter, $\varepsilon_F$ the Fermi energy imposed by the electrodes and $\theta$ the temperature. Solving the percolation problem, VM found the conductance of the organic thin film [R1]

$$G = A_0 \exp(\beta \varepsilon_F) \qquad (S3)$$

where $\beta = 1/ k_B \theta$ and $A_0$ is a dimensionless function of $N_t$, $\theta_0$, $\theta$ and $\alpha$. The charge carrier density at the organic semiconductor/dielectric interface is given by the following integral

$$N_P = \int_{-\infty}^{+\infty} d\varepsilon \, D(\varepsilon) f(\varepsilon, \varepsilon_F) \approx B_0 N_t \exp(\beta_0 \varepsilon_F) \qquad (S4)$$

where $f$ is the Fermi-Dirac distribution, $\beta = 1/ k_B \theta$, and $B_0$ is a function with no dimension of $\theta$ and $\theta_0$. The last equality is valid for low carrier density [R1].

When a gate potential is applied, an accumulation layer is formed at the interface between the film and the dielectric as discussed above (eq. [S4]), but charges are also stored in the NPs. We note $Q_P(t)$ and $Q_{NP}(t)$, the number of holes in the accumulation layer and in the NPs, respectively. $Q_P(t)=\Omega N_P(t)/e$, $\Omega$ is the volume of the thin film, $e$ the elementary charge. We write $Q_T(t)= Q_P(t) + Q_{NP}(t)$, the total number of holes. The charges $Q_P(t)$ and $Q_{NP}(t)$ interact via Coulomb interaction changing the site energies of the electrical network, $\varepsilon$. These modifications were modelled in Ref.





[R2] by a shift, up or down, depending on the sign of the charges, of the Fermi level by an amount of $-\Delta$. We wrote

$$G = A_0 \exp(\beta \varepsilon_F - \beta \Delta) \tag{S5}$$

for the conductance. In the same way, we can write

$$Q_P = \frac{\Omega}{e} \int_{-\infty}^{+\infty} d\varepsilon \, D(\varepsilon) f(\varepsilon, \varepsilon_F - \Delta) \approx \frac{\Omega}{e} \exp(-\beta \Delta) \int_{-\infty}^{+\infty} d\varepsilon \, D(\varepsilon) f(\varepsilon, \varepsilon_F) \tag{S6}$$

for the number of holes in the accumulation layer. The second equality is valid if $\beta \Delta << 1$. Combining Eqs. (S4) and (S6), we get

$$Q_P \approx \frac{\Omega}{e} B_0 N_t \exp(\beta_0 \varepsilon_F - \beta \Delta) \tag{S7}$$

Starting from Eqs. (S6) and (S7) we show in the following that the NOMFET is a memristive device as defined by Chua [R4,R5]. Since the NOMFET is used as a pseudo two terminal device, there is only one control parameter, $V_{DS}$, the potential applied between the source and gate/drain (these later connected together). When $V_{DS}$ is applied, both $Q_P(t)$ and $Q_{NP}(t)$ are changed. As a consequence, at time $t$, the drain-source current, $I_{DS}$, can be written as usual for memristive devices [R4,R5]

$$I_{DS}(t) = \sigma_0(V_{DS}) f(Q_{NP}(t)) V_{DS}(t) \tag{S8}$$

$\sigma_0$ is the conductivity of our device that includes the field effect. $f(Q_{NP})$ is the memristive function of the device that contains the effects of the charges trapped in the NPs [R4,R5]. $Q_{NP}(t)$ is the so-called internal parameter of the NOMFET. Based on our previous analysis, we write

$$f(Q_{NP}(t)) = e^{-\gamma Q_{NP}(t)} \tag{S9}$$

Comparing with Eq. (S5) we identify $\sigma_0(V_{DS}) = A_0 \exp(\beta \varepsilon_F(V_{DS}))$ and $\gamma Q_{NP} = \Delta \beta$. We assume in the following that the $V_{DS}$ dependence in the second expression is uniquely due to $Q_{NP}$ and consider $\gamma$ as a constant. We note that $1/\gamma$ behaves as a capacitance times $k_B \theta$. We next simplify further Eq. (S8) considering the effect of the trapped charges as a perturbation: we write

$$I_{DS}(t) \approx \sigma_0(V_{DS})(1 - \gamma Q_{NP}(t)) V_{DS}(t) \tag{S10}$$

In the same way, starting from Eq. (S7) we do the following series expansion

$$Q_P(t) \approx \frac{\Omega}{e} B_0 N_t \left(1 - \beta_0 \varepsilon_F - \gamma Q_{NP}(t)\right) \tag{S11}$$





Finally, we rewrite the second term showing explicitly the $V_{DS}$ dependence

$$Q_P(t) \approx \overline{N} - \eta(V_{DS})V_{DS}(t) - \gamma\overline{N}Q_{NP}(t) \tag{S12}$$

where $\overline{N} = \Omega B_0 N_t / e$ and $\eta$ is a function of $V_{DS}$ to be fitted on experiments. The term $\eta(V_{DS})V_{DS}$ models the Fermi level shift. The carrier density depends on the gate potential, as usual, but also on $Q_{NP}$ that shifts the Fermi level.

This model was already used to describe the facilitating and depressing synaptic behaviours of our device [R2]. It gives very good agreements with experiments but all the measurements were done at fixed $V_{DS}$. In the present work we need to consider explicitly the $V_{DS}$ dependence. Moreover, the time trajectory of $I_{DS}$ depends on the details of the charge/discharge dynamics of the nanoparticles that needs to be specified further.

## 2- *Dynamics of charge/discharge of the gold nanoparticle.*

We choose to describe the charge/discharge dynamics of the NPs by the simplest as possible kinetic equations

$$\begin{cases} \dot{Q}_{NP} = -k_{NP\to P}(V_{DS})Q_{NP} + k_{P\to NP}(V_{DS})Q_P \\ \dot{Q}_P = -k_{P\to NP}(V_{DS})Q_P + k_{NP\to P}(V_{DS})Q_{NP} + SQ_P(V_{DS}) \end{cases} \tag{S13}$$

The dot is for the time derivative. The rate coefficients, $k_{P\to NP}(V_{DS})$ - NP charging, and $k_{NP\to P}(V_{DS})$ - NP discharging, depend on the control parameter $V_{DS}$. The number of charges in the pentacene film is not constant but depends on $V_{DS}$ and on the number of charges trapped in the NPs, $Q_{NP}$. $SQ_P$ is the source term that gives the amount of positive charges created at time $t$ coming from source and drain electrodes. From Eq. (S12) we can deduce

$$SQ_P(V_{DS}) = -\eta(V_{DS})\dot{V}_{DS} - \gamma\overline{N}\dot{Q}_{NP} \tag{S14}$$

At fixed $V_{DS}$, once stationarity is reached the charge variables fulfil

$$\begin{cases} 0 = -k_{NP\to P}(V_{DS})Q_{NP}^{st} + k_{P\to NP}(V_{DS})Q_P^{st} \\ 0 = -k_{P\to NP}(V_{DS})Q_P^{st} + k_{NP\to P}(V_{DS})Q_{NP}^{st} \end{cases} \tag{S15}$$

with the condition

$$Q_T(t) = Q_T^{st} = Q_{NP}^{st} + Q_P^{st} \tag{S16}$$





meaning that the number of holes and their repartition between pentacene and particles are time independent. They however depend on $V_{DS}$. We get

$$Q_{NP}^{st}(V_{DS}) = \frac{k_{P \to NP}(V_{DS})}{k_{NP \to P}(V_{DS}) + k_{P \to NP}(V_{DS})} Q_T^{st}(V_{DS}) \tag{S17}$$

To characterize efficiently our device we systematically apply the same relaxation procedure described in the main text before any sequences of measurements. We assume then the device to be at equilibrium (or at rest). This state is characterized by $Q_{NP}^0$, $Q_P^0$ and $Q_T^0$ defined as

$Q_{NP}^0 = Q_{NP}^{st}(0)$, $Q_P^0 = Q_P^{st}(0)$ and $Q_T^0 = Q_T^{st}(0)$. We then measure the variation of charges trapped in the NPs, $\delta Q_{NP}$, with respect to this reference state after various types of excitations described in the main text and below. This gives us a rational way to characterize our device.

If we suddenly switch on $V_{DS}$ just after the relaxation step the current reads

$$I_0 \approx \sigma_0(V_{DS})(1 - \gamma Q_{NP}^0)V_{DS} \tag{S18}$$

This is the reference current at voltage $V_{DS}$. Out of stationarity, we write Eqs (S13) as

$$\dot{Q}_{NP} = -\left(k_{NP \to P}(V_{DS}) + k_{P \to NP}(V_{DS})\right)Q_{NP} + k_{P \to NP}(V_{DS})Q_T(t) \tag{S19}$$

with

$$Q_T(t) = Q_T^0 + \delta Q_T(t) = Q_P^0 - \gamma \overline{N}(Q_{NP}(t) - Q_{NP}^0) - \eta(V_{DS})V_{DS}(t) \tag{S20}$$

We then obtain a simple first order differential equation for $Q_{NP}$

$$\dot{Q}_{NP} = -\frac{1}{\tau(V_{DS})}Q_{NP} + \frac{1}{\tau_0(V_{DS})}Q_{NP}^0 - \eta(V_{DS})V_{DS}(t) \tag{S21}$$

where the characteristic time constants are

$$\begin{cases} \dfrac{1}{\tau(V_{DS})} = k_{P \to NP}(V_{DS}) + k_{NP \to P}(V_{DS}) + \gamma \overline{N} k_{P \to NP}(V_{DS}) \\ \dfrac{1}{\tau_0(V_{DS})} = k_{P \to NP}(V_{DS})\left(\dfrac{k_{P \to NP}(0) + k_{NP \to P}(0)}{k_{P \to NP}(0)} + \gamma \overline{N}\right) \end{cases} \tag{S22}$$

At $V_{DS}$=0, starting from an initial charge, Eq. (S21) gives an exponential relaxation. We have seen in Ref. [R2] that the charge relaxations of our device follow power laws that can be in general approximated by a single exponential but in a limited time interval. It is therefore clear that our





simple set of equations (Eqs. (S21) and (S22)) could only work in such limited time interval. For larger time window of observation, more sophisticated theory would be needed. It turns out that this simple modelling is sufficient for our purpose.

The complete determination of the rate coefficients implies a microscopic study of the hole tunnelling between pentacene and nanoparticles. This is a complex problem that goes far beyond the scope of the present study. Instead we apply series of reasonable approximations detailed below.

The first approximation concern the source term (Eq. (S14)) that is a key ingredient of the kinetic equation (Eqs. (S13)). It characterizes the amount of holes created at time $t$ in the accumulation layer of the pentacene. The NOMFET, as all the organic thin film transistors, is not bipolar. It means that accumulations of holes or electrons are not equivalent and obey different properties: it is more difficult to accumulate electrons than holes. We decompose the source term in two different components: one for negative voltage (accumulation of holes) and one for the positive voltage (depletion of holes). Guided by our experimental results (see Fig. 2-c in the main text) – as it will be clear below - we assume that the depletion of holes becomes efficient only for voltages higher than a threshold, $V_{th}$. We write

$$SQ_P(V_{DS}) = -\left[\eta_-(V_{DS})\Theta(-V_{DS}) + \eta_+(V_{DS})\Theta(V_{DS} - V_{th})\right]\dot{V}_{DS} - \gamma\overline{N}\dot{Q}_{NP}$$  (S23)

introducing two new functions $\eta_-$ and $\eta_+$ for the negative and positive branches, respectively. $\Theta$ is the Heaviside function. Doing so we divide the voltage space in three regions: $V_{DS} < 0$, $0 < V_{DS} < V_{th}$ and $V_{DS} > V_{th}$.

The second set of approximations concern the $V_{DS}$ dependence of the rate coefficients. It is detailed in the next section.

### 3- *Comparisons with experiments.*

To characterize our device, starting from the reference point we apply sequences of pulses and consider

$$\frac{I_{DS}(t) - I_0}{I_0} = e^{-\gamma\delta Q_{NP}(t)} - 1 \approx -\gamma\delta Q_{NP}(t) = -\gamma(Q_{NP}(t) - Q_{NP}^0)$$  (S24)

which is approximately proportional to the changes of charges trapped in the NPs at time $t$ after a particular history of the external parameter $V_{DS}$. Note that we assume that the trapped charges are small enough to have $\gamma\delta Q_{NP} \ll 1$. This will be check on the experiments (see main text, section 3). We





have applied three different types of signal: A single pulse (Fig. 2-b, main text) and two different sequences of spikes that differ by the shape of those spikes (Spike 1 - rectangular shape and Spike 2 – triangular shape, see Fig. 3-a, main text). In the following we detail the results that we obtain with our model to describe each of these experiments.

<u>Single pulse experiments</u>. This type of experiment gives an easy way to characterize our devices through the variations of the internal parameter. The current depends on three parameters: the amplitude, $V_{DS} = V_P$, and the width, $W$, of the pulse and on the final time, $t$, of the experiment where the final current is measured. We get

$$\frac{I(V_{DS},W,t) - I_0}{I_0} = e^{-\gamma \delta Q_{NP}(V,W,t)} - 1 \approx -\gamma \delta Q_{NP}(V_{DS},W,t) = -\gamma(Q_{NP}(V_{DS},W,t) - Q_{NP}^0) \qquad (S25)$$

that can be written as

$$-\gamma \delta Q_{NP}(V_{DS},W,t) = \gamma \left[ \tau(V_{DS})\eta(V_{DS})V_{DS} + Q_{NP}^0 \left(1 - \frac{\tau(V_{DS})}{\tau_0(V_{DS})}\right) \right] \left(1 - \exp\left(-\frac{W}{\tau(V_{DS})}\right)\right) \exp\left(-\frac{(t-W)}{\tau(0)}\right) \qquad (S26)$$

To go further, we need to specify the different functions of $V_{DS}$ appearing in the above expression. In a first approximation (Approximation 1), we simply neglect the voltage dependence in each region of $V_{DS}$ of the rate coefficients. This could be justified by the fact that the particles and the pentacene molecules involved in the tunnelling process are very close in space and so at approximately the same potential. We write

$$k_{\substack{NP \to P \\ P \to NP}}(V_{DS}) = \begin{cases} k_{\substack{NP \to P \\ P \to NP}}^-, V_{DS} < 0 \\ k_{\substack{NP \to P \\ P \to NP}}^0, 0 < V_{DS} < V_{th} \\ k_{\substack{NP \to P \\ P \to NP}}^+, V_{DS} > V_{th} \end{cases} \qquad (S27)$$

To reproduce the data in the simplest way, the best strategy is to consider together with the above rate coefficients, the following source term

$$SQ_P = -\left(\eta_- \Theta(-V_{DS}) + \eta_+ \Theta(V_{DS} - V_{th})\right)\dot{V}_{DS} - \gamma \bar{N} \dot{Q}_{NP} \qquad (S28)$$

where $\eta_-$ and $\eta_+$ are constants to be determined. We must then assume also

$$\tau(V_{DS}) = \tau_0(V_{DS}) \qquad (S29)$$





to avoid any discontinuity as function of $V_{DS}$ in the internal parameter. These approximations are equivalent to consider three different time constants

$$\tau(V_{DS}) = \begin{cases} \tau_-, V_{DS} < 0 \\ \tau_0, 0 < V_{DS} < V_{th} \\ \tau_+, V_{DS} > V_{th} \end{cases} \tag{S30}$$

The internal parameter takes then a particularly simple form

$$-\gamma \delta Q_{NP}(V_{DS}, W, t) = \begin{cases} \gamma \tau_- \eta_- V_{DS}\left(1 - \exp\left(-\dfrac{W}{\tau_-}\right)\right)\exp\left(-\dfrac{(t-W)}{\tau_0}\right), & V_{DS} < 0 \\ 0, \quad 0 < V_{DS} < V_{th} \\ \gamma \tau_+ \eta_+ (V_{DS} - V_{th})\left(1 - \exp\left(-\dfrac{W}{\tau_+}\right)\right)\exp\left(-\dfrac{(t-W)}{\tau_0}\right), & V_{DS} > V_{th} \end{cases} \tag{S31}$$

It gives two linear relations: one starting at $V_{DS}$=0 and the other at $V_{DS}$=$V_{th}$ that fit relatively well our data (see Fig. 2-c, main text). There are six parameters to optimize: $V_{th}$, $\tau_0$, $\tau_-$, $\tau_+$, $\gamma\eta_-$ and $\gamma\eta_+$. It is obvious that an infinite set of parameter values give the same proposed function. However $V_{th}$ is unambiguously fixed by the shape of the memristive function to $V_{th}$=15V. We choose to fix the characteristic time constants to reasonable values taken from experiments [1, 3] (typically $\tau_0$, $\tau_-$, $\tau_+$ ≈ 1-5s); the two remaining parameters are then uniquely determined, to fit the experimental curve (Fig. 2-c) to $\gamma\tau_-\eta_-$=1.7x10$^{-2}$ and $\gamma\tau_+\eta_+$=1.3x10$^{-2}$.

<u>Spike 1 and Spike 2</u>. Different sequences of pre- and post- synaptic spikes distant by a certain time interval $\Delta t$ are applied to the NOMFET. They allow us to evidence the STDP properties of our device. Two kind of spike are used that give different outcomes (Figs. 4-a and 4-b, main text). To get the corresponding STDP functions (Eq. (S24)), one has to solve the first order differential equation (S21) with the appropriate $V_{DS}(t)$: to each sequence of spikes with a given $\Delta t$ corresponds a particular $V_{DS}(t)$. We have then, as for the previous case, to specify the different functions of $V_{DS}$ that appear in Eq. (S21). Applying the simple Approximation 1 does not give good results especially for the spike 2. For spike 1 (rectangular), the tendency could be reproduced by choosing very long characteristic times ($\tau_0$, $\tau_-$, $\tau_+$ ≈ 1000s) but the amplitude of the response is then by far too low; for spike 2 (triangular), the response is totally inconsistent with the data. In a second approximation (Approximation 2), we do not consider the rate coefficients as simple constants in each voltage region (Eq. (S27)) but instead recognize that the voltage dependences could become important





above thresholds: $V_-$ and $V_+$ in the negative and positive region, respectively. These dependences are taken in the simplest way

$$k_{\substack{NP \to P \\ P \to NP}}(V_{DS}) = \begin{cases} k_{\substack{NP \to P \\ P \to NP}}^{-2}, V_{DS} < V_- \\ k_{\substack{NP \to P \\ P \to NP}}^{-1}, V_- < V_{DS} < 0 \\ k_{\substack{NP \to P \\ P \to NP}}^{0}, 0 < V_{DS} < V_{th} \\ k_{\substack{NP \to P \\ P \to NP}}^{+1}, V_+ > V_{DS} > V_{th} \\ k_{\substack{NP \to P \\ P \to NP}}^{+2}, V_{DS} > V_+ \end{cases} \qquad (S32)$$

We consider with Eq. (S32) the corresponding source term

$$SQ_P = -\left((\eta_{-2} - \eta_{-1})\Theta(-V_{DS} + V_-) + \eta_{-1}\Theta(-V_{DS}) + \eta_{+1}\Theta(V_{DS} - V_{th}) + (\eta_{+2} - \eta_{+1})\Theta(V_{DS} - V_+)\right)\dot{V}_{DS} - \gamma\dot{Q}_{NP} \qquad (S33)$$

These approximations are equivalent to consider five different time constants

$$\tau(V_{DS}) = \begin{cases} \tau_{-2}, V_{DS} < V_- \\ \tau_{-1}, V_- < V_{DS} < 0 \\ \tau_{0}, 0 < V_{DS} < V_{th} \\ \tau_{+1}, V_+ > V_{DS} > V_{th} \\ \tau_{+2}, V_{DS} > V_+ \end{cases} \qquad (S34)$$

With this new set of approximation, the differential equation (S21) is decomposed in different pieces depending on the $V_{DS}$ amplitude, according to Eq. (S34), and then solved. We impose Approximation 2 to give the same result for the memristive function (Fig. 2-c, main text). During the fitting procedure, once the time characteristics are chosen the different $\eta$ constants are fixed by imposing the linear relations shown in Fig. 2-c (main text). Choosing $\tau_0$, $\tau_{-1}$, $\tau_{+1}$ = 5s, $\tau_{-2}$ = 1s and $\tau_{+2}$ = 0.3s for the characteristics time constants and $V_-$ = -25V and $V_+$ = 40V give the best fits shown in Figs. 4-a and 4-b (main text). With the same collection of parameters, the whole set of data is reasonably well fitted by Approximation 2.





**Reset Protocol.**

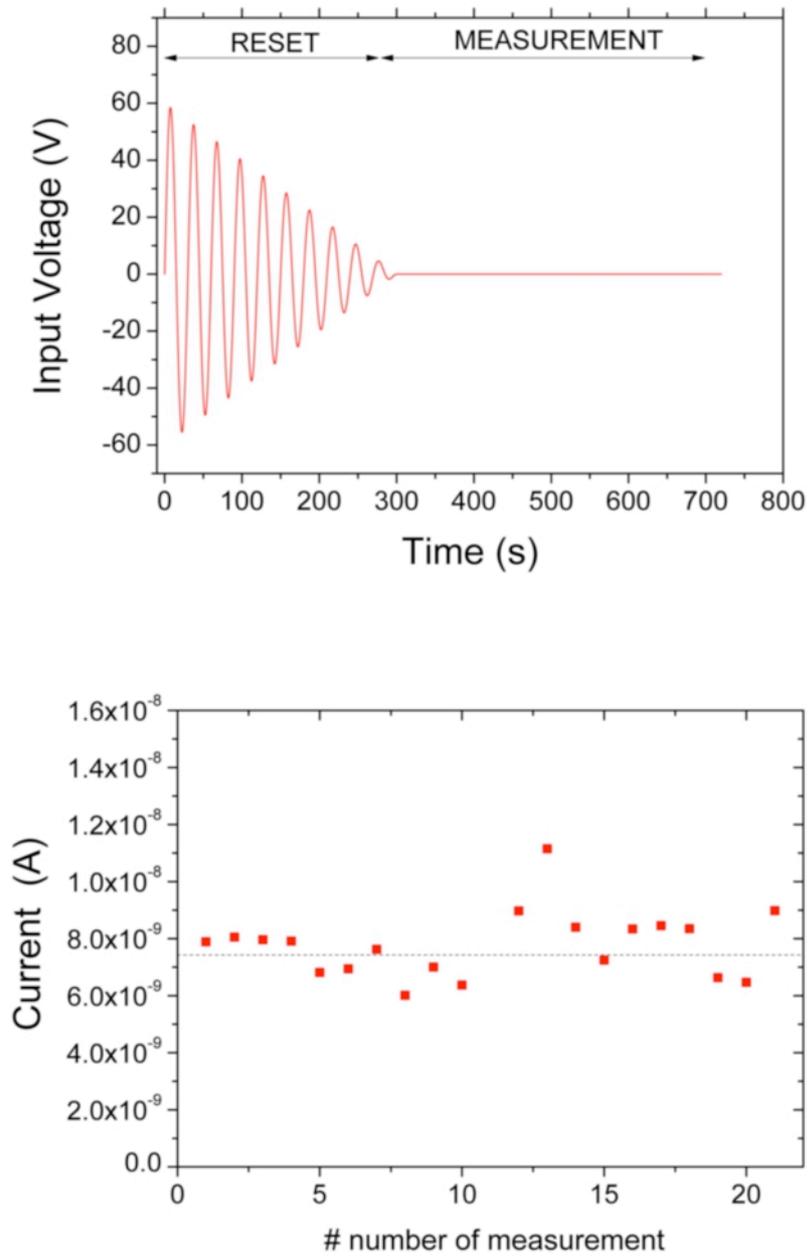

***Figure S1. (a)*** *Typical reset signal applied to the NOMFET to reset the same intial state (i.e. almost the same charge in the NPs and same output drain current) before each new SDTP measurements.* ***(b)*** *Typical drain current measured after the reset signal for more than 20 experiments.*





**Physical meaning of threshold voltages in curve 2-c (memristive function)**

For $V_{DS} < V_{th1}$, when the NOMFET is in its ON state, the NP charging mechanism is likely hole tunneling through the organic capping layer of the NP (typically alkylthiol, see Experimental section). This tunnel trapping can start as soon as holes are present in the OSC at the interface, since the turn-on voltage of the NOMFET is around 0 V (albeit with a large device-to-device dispersion, ± 5 V), it explains $V_{th1} \approx 0$ V. The slope of this part is about $1.7 \times 10^{-2}$ (or 1/59) V$^{-1}$. For $V > V_{th2}$, the OSC is in depletion, and the hole detrapping occurs probably through field-assisted emission. In that case, $V_{th2}$ relies on the minimum internal field required to overcome the energy barrier at the NP/OSC interface for charge detrapping. The slope of this part is about $1.3 \times 10^{-2}$ (or 1/79) V$^{-1}$. The difference in the slopes would indicate that tunneling trapping is more efficient than field-assisted detrapping in the present case. A detailed analysis of these charging/discharging phenomena in the NOMFET is beyond the scope of this paper and will require more voltage-dependent and temperature-dependent experiments.

**Electronic board for hybrid CMOS/NOMFET measurements.**

Basically, the electronic board comprises three essential parts. 1) The pulse (Fig. S2-b) are generated with an analog multiplexer (MAX14752) which can wwitch voltages up to ± 36V). 2) The current is measured on board with an op-amp current-to-voltage converter (OPA445) supporting a large voltage up to ± 45V and having a low input bias current (ca. 10 pA), required to measure current down to the nA. It is followed by an analog-to-digital converter (ADC, model LTC1856) to obtain a digital value of the current. 3) This electronic board is fully controlled with a FPGA (Field Programmable Gate Array) board, which is driven by a PC.

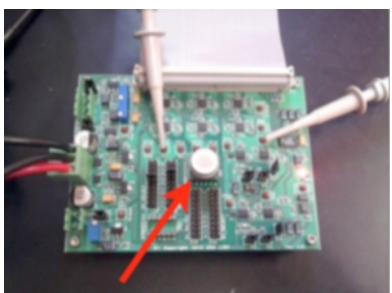
(a)

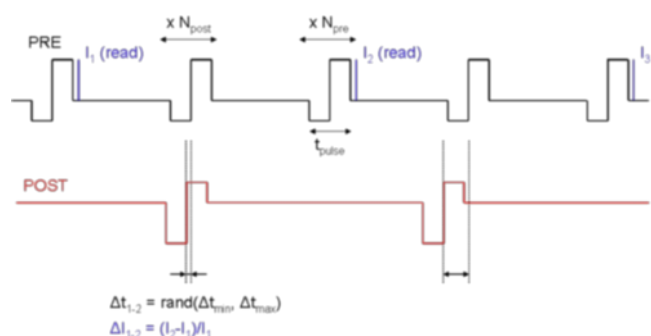
(b)

**Figure S2.** *(a) Photography of the NOMFETs in a TO case (arrow) plugged on the electronic-board. (b) Typical spike signals applied at the input (pre-synaptic spikes) and at the output (pots-synaptic spike)*





*of the NOMFET for measing data shown in Fig. 4-c. For each Δt, a first sequence of $N_{pre}$ = 20 pre-synaptic spikes are applied before the conductivity is measured, followed by a sequence of $N_{post}$ = 20 pre-post spike interactions, after which the conductivity is measured again and compared to the previous one to obtain the change of conductivity. The pre-synaptic spike's duration is 4 s (-15 V during 2 s followed by 30 V during 2 s), as the post-synaptic spike (-30 V during 2 s followed by 15 V during 2 s).*

**Behavioral model equations.**

From Eqs. 1-2 and S9, S10, S21, S24, we can rewrite Eq. 1 and 2 as:

$$I_{DS}(t) = \sigma_0(V_{DS}(t))V_{DS}(t)e^{-\gamma Q_{NP}(t)}$$
$$\gamma\tau\dot{Q}_{NP}(t) = -\gamma Q_{NP}(t) - \gamma\tau\eta(V_{DS}(t))V_{DS}(t)$$

(S35)

For macro-modeling convenience, let us define a circuit variable (a voltage) *w(t)* to describe the evolution of NP charges in the NOMFET:

$$w(t)\big/w_0 = \gamma Q_{NP}(t)$$

(S36)

where $w_o$ is a normalization constant that takes the value of 1*V*. Similarly, let us redefine the second right hand side term of bottom equation (S35) as a single function of the NOMFET terminal voltages as:

$$\rho(V_{DS}(t)) = \gamma w_0\eta(V_{DS}(t))V_{DS}(t)$$

(S37)

As a result, eq. (S35) becomes

$$I_{DS} = \sigma_0(V_{DS})V_{DS}e^{-w/w_0}$$
$$\frac{\tau\dot{w}}{w_0} = -\frac{w}{w_0} - \frac{\tau}{w_0}\rho(V_{DS})$$

(S38)

The function *ρ(V_DS)* can be described by (see S31):

$$\frac{\tau}{w_0}\rho(V_{DS}) = \gamma\tau\eta(V_{DS})V_{SD} = \begin{cases} (V_{DS} - V_{th})\gamma\tau\eta^+ &, \quad V_{DS} > V_{th} \\ 0 &, \quad 0 < V_{SD} < V_{th} \\ V_{DS}\gamma\tau\eta^- &, \quad V_{DS} < 0 \end{cases}$$

(S39)

with $V_{th}$ = 15 V, $\gamma\tau\eta^+ \approx 0.013$ V$^{-1}$ and $\gamma\tau\eta^- \approx 0.017$ V$^{-1}$ as given after fitting Eq. (S31) to data of Fig. 2-c.

**REFERENCES**


[1]     F. Alibart, *et al.*, "An Organic Nanoparticle Transistor Behaving as a Biological Spiking Synapse," *Advanced Functional Materials,* vol. 20, pp. 330-337, 2010.







[2]     M. C. J. M. Vissenberg and M. Matters, "Theory of the field-effect mobility in amorphous organic transistors," *Phys. Rev. B,* vol. 57, pp. 12964-12967, 1998.

[3]     O. Bichler, *et al.*, "Functional Model of a Nanoparticle-Organic Memory Transistor for Use as a Spiking Synapse," *IEEE Trans. Electron Devices,* vol. 57, pp. 3115-3122, 2010.

[4]     L. O. Chua, "Memristor - the missing circuit element," *IEEE Trans. on Circuit Theory,* vol. 18, pp. 507-519, 1971.

[5]     L. O. Chua and S. M. Kang, "Memristive devices and systems," *Proc. of the IEEE,* vol. 64, pp. 209-223, 1976.